\documentclass[12pt]{article}

\begin{document}

\title{
{\Large \bf Construction of a non-standard quantum field theory
through a generalized Heisenberg algebra}
}         

\author{
 M. A. Rego-Monteiro$^{a}$ and E. M. F. Curado$^{b}$  \\
Centro Brasileiro de Pesquisas F\'\i sicas, \\ 
Rua Xavier Sigaud 150, 22290-180 - Rio de Janeiro, RJ, Brazil\\
$^a$regomont@cbpf.br \\
$^b$eme@cbpf.br
}        
\date{}          
\maketitle
\begin{abstract}
 
\indent

	We construct a Heisenberg-like algebra for the one dimensional
quantum free Klein-Gordon equation defined on the interval of the
real line of length $L$. Using the realization of the ladder
operators of this type Heisenberg algebra in terms of physical 
operators we build a $3+1$ dimensional free quantum field theory 
based on this algebra. 
We introduce fields written in terms of the ladder operators
of this type Heisenberg algebra and a free quantum Hamiltonian
in terms of these fields. The mass spectrum of the physical 
excitations of this quantum field theory are given by 
$\sqrt{n^2 \pi^2/L^2+m_q^2}$, where $n= 1,2,\cdots$ denotes 
the level of the particle with mass $m_q$ in an infinite square-well 
potential of width $L$.

\end{abstract}

\vspace{1cm}

\begin{tabbing}

\=xxxxxxxxxxxxxxxxxx\= \kill

\\  $q$-oscillators; hadrons.


\end{tabbing}

\newpage

\section{Introduction}

There is a vast range of energy from the 
present accelerators energies ($\approx 10^3$Gev)
to the Planck energy
($10^{19}$ Gev)  where it is believed
that there is room for surprises $^{\cite{wilczek}}$. 
It is also believed that field theories
based on deformed algebras could play an important
role to describe physics in this vast energy range 
$^{\cite{qalg}}$. These algebras have parameters,
known as deformation parameters, that it is expected 
to regularize the ultraviolet divergences in 
deformed field theories $^{\cite{castella}}$.

This paper is the first step towards
an analysis of possible consequences in
quantum field theories (QFTs) of a class of 
generalized Heisenberg algebras we have recently
constructed $^{\cite{jpa}}$. The generators of each 
algebra in this class are the Hamiltonian of the 
one-dimensional quantum system in consideration
and the ladder operators. The physical systems
that are described by these type Heisenberg 
algebras are characterized by those one-dimensional
quantum systems having an spectrum where the successive 
energy levels are related by $\epsilon_{n+1}=f(\epsilon_n)$
$^{\cite{comhugo}}$.

Within this class of algebras we find deformed
and also non-deformed type Heisenberg algebras 
$^{\cite{jpa}}$ and as
a first step of this program we explore the 
consequences in QFT
of a non-deformed type Heisenberg algebra belonging to
the large above mentioned class. We hope that the QFT 
we obtain in this paper 
can have an intrinsic interest as an alternative 
phenomenological QFT for hadronic interactions.
Moreover, the whole procedure could be seen as a 
prototype since it seems possible
to implement the approach developed here
to a deformed Heisenberg algebra belonging to the
above mentioned class of algebras in order to construct
a deformed QFT that could be appropriate for very high energy 
($10^3$Gev$ < E < 10^{19}$Gev) physics.

Our approach extends an standard aspect of QFTs. 
In the
very well-known theory of quantum spin-$0$ particles
$^{\cite{tdlee}}$,
scalar particle states appear in QFT as vacuum excitations
through the application of the Heisenberg algebra
creation operator to the vacuum. Roughly speaking,
what we do here is to construct a free QFT where the
Hamiltonian eigenvectors are obtained by the successive 
application to the vacuum of the creation operator 
of another physical system instead the ordinary 
harmonic oscillator.

We construct here the Heisenberg-like algebra of a 
relativistic particle in an infinite square-well potential: 
the quantum one dimensional free Klein-Gordon equation 
defined on the interval $0 \leq x \leq L$ with special 
boundary conditions. 
Within this picture, the creation
operator of the algebra when applied to the vacuum of the theory
creates particle states  with mass spectrum 
$\sqrt{n^2 \pi^2/L^2+m_q^2}$ where $n = 1,2,\cdots$
gives the level of the particle with mass $m_q$ in an infinite
square-well potential of width $L$. Afterwards, we use the 
ladder operators of this physical system to construct fields 
and a free QFT Hamiltonian. 

In section 2, we briefly review a class of generalized Heisenberg algebras
having $q$-oscillators as a particular example of this class. Within
the general class of Heisenberg algebras presented in section 2 there is also
the Heisenberg algebra of a relativistic particle in an one-dimensional
infinite square-well potential (a non-deformed type Heisenberg algebra). 
This relativistic square-well algebra
is presented in section 3 where it is also given the realization of the
ladder operators in terms of the physical operators of the model. In section
4, we construct the first steps towards a quantum field theory based on
the square-well Heisenberg algebra. We introduce in this section 
fields written in terms of the generators of the algebra
and a Hamiltonian written in terms of these fields, describing
a system with an infinite number of degrees of freedom 
with mass spectrum excitations given by 
$\sqrt{n^2 \pi^2/L^2+m_q^2}$,
where $n = 1,2,\cdots$
denotes the level of the particle with mass $m_q$ in an infinite
square-well potential of width $L$. In section 5 we present our
final comments where we conjecture a possible application of this
formalism as an alternative phenomenological quantum field 
approach to hadronic interactions.

\section{Generalized Heisenberg algebras}

Let us consider an algebra generated by $J_{0}$, $A$ and $A^{\dagger}$ described
by the relations $^{\cite{jpa}}$
\begin{eqnarray}
J_{0} \, A^{\dagger} &=& A^{\dagger} \, f(J_{0}) ,
\label{eq:alg1} \\
A \, J_{0} &=& f(J_{0}) \, A , 
\label{eq:alg2} \\
\left[ A^{\dagger},A \right] &=& J_{0}-f(J_{0}) ,
\label{eq:alg3}
\end{eqnarray}
where $^{\dagger}$ is the Hermitian conjugate and, 
by hypothesis, $J_{0}^{\dagger}=J_{0}$ and $f(J_{0})$
is a general analytic function of $J_{0}$. It is easy to see that the 
Jacobi identity is trivially satisfied for general $f$.
The case where 
$f(J_{0})=r \, J_{0} \, (1-J_{0})$ was analyzed in ref. \cite{algebra1}. 

Using the algebraic relations in eqs. (\ref{eq:alg1}-\ref{eq:alg3}) we
see that the operator
\begin{equation}
C = A^{\dagger} \, A - J_{0} = A \, A^{\dagger} - f(J_{0})  
\label{eq:casimir}
\end{equation}
satisfies
\begin{equation}
\left[ C,J_{0} \right] = \left[ C,A \right] = 
\left[ C,A^{\dagger} \right] = 0  ,
\label{eq:comute}
\end{equation}
being thus a Casimir operator of the algebra.

We analyze now the representation theory of the algebra when 
the function $f(J_{0})$ is a general analytic function of  
$J_{0}$. 
We assume we have an $n$-dimensional irreducible representation
of the algebra given in eqs. (\ref{eq:alg1}-\ref{eq:alg3}). 
Consider the state $|0\rangle$ with the lowest 
eigenvalue of the Hermitian operator $J_{0}$
\begin{equation}
J_{0} \, |0\rangle = \alpha_{0} \, |0\rangle .
\label{eq:alfa0}
\end{equation}
For each value of $\alpha_{0}$ and the parameters of the algebra
we have a different vacuum that for simplicity will be denoted by
$|0\rangle$.  

Let $| m \rangle$ be a normalized eigenstate of $J_{0}$, 
\begin{equation}
    J_{0} |m \rangle = \alpha_{m} |m \rangle \, . 
    \label{eq:alfam}
\end{equation}
Applying eq. (\ref{eq:alg1}) to $|m \rangle$ we have  
\begin{equation}
    J_{0} (A^{\dagger} |m \rangle) = A^{\dagger} f(J_{0}) |m \rangle = 
    f(\alpha_{m}) (A^{\dagger} |m \rangle ) \, .
    \label{eq:j+}
\end{equation} 
Thus, we see that $A^{\dagger} |m \rangle $ is a $J_{0}$ 
eigenvector with eigenvalue $f(\alpha_{m})$.  
Starting from $|0 \rangle$ and applying successively $A^{\dagger}$ to 
$|0 \rangle$ we create different states with $J_{0}$ eigenvalue 
given by 
\begin{equation}
    J_{0} \left( (A^{\dagger})^m |0 \rangle \right) = 
    f^m (\alpha_{0}) \left( (A^{\dagger})^m |0 \rangle \right) \, ,
    \label{eq:j+m}
\end{equation}
where $f^m (\alpha_{0})$ denotes the $m$-th iterate of $f$.  Since 
the application of $A^{\dagger}$ creates a new vector, whose respective 
$J_{0}$ eigenvalue has iterations of $\alpha_{0}$ through $f$ 
augmented by one unit, it is 
convenient to define the new vectors $(A^{\dagger})^m |0 \rangle$ as 
proportional to $|m \rangle$ and we then call $A^{\dagger}$ a raising 
operator.  Note that  
\begin{equation}
\alpha_m = f^m(\alpha_0) = f(\alpha_{m-1}) \, ,
    \label{eq:alfam3}
\end{equation}
where $m$ denotes the number of iterations of $\alpha_{0}$ 
through $f$.  

Following the same procedure for $A$,  applying eq. (\ref{eq:alg2}) 
to $|m+1 \rangle$, we have 
\begin{equation}
    A \, J_{0} |m+1 \rangle = f(J_{0}) \left( A |m+1 \rangle \right) =
    \alpha_{m+1} \left( A |m+1 \rangle \right) \, ,
    \label{eq:alfam2}
\end{equation}
showing that $A \, |m+1 \rangle$ is also a $J_{0}$ eigenvector with 
eigenvalue $\alpha_{m}$.  Then, $A \, |m+1 \rangle $ is proportional 
to $|m \rangle$ being $A$ a lowering operator.  

Since we consider 
$\alpha_{0}$ the lowest $J_{0}$ eigenvalue, we require
\begin{equation}
A \, |0\rangle = 0 .
\label{eq:vacuum}
\end{equation}
As was shown in \cite{algebra1}, depending on the function $f$ 
and its initial value $\alpha_{0}$, it may happen that the 
$J_{0}$ eigenvalue of state $|m+1 \rangle$ is lower than the one 
of state $|m \rangle$.   Then, as shown in \cite{jpa}, 
given an arbitrary analytical function $f$ 
(and its associated algebra in eqs. (\ref{eq:alg1}-\ref{eq:alg3})) 
in order to satisfy eq. (\ref{eq:vacuum}),   
the allowed values of $\alpha_{0}$ are chosen in such a way that the  
iterations $f^m (\alpha_{0})$ ($m \geq 1$) are 
always bigger than $\alpha_{0}$.

As was proven in \cite{jpa} in general we obtain
\begin{eqnarray}
J_{0} \, |m\rangle &=& f^{m}(\alpha_0) \, |m\rangle , \; \; \; m = 0,1,2, 
\cdots \; , 
\label{eq:b1} \\
A^{\dagger} \, |m-1\rangle &=& N_{m-1} \, |m\rangle , 
\label{eq:b2} \\
A \, |m\rangle &=& N_{m-1} \, |m-1\rangle ,
\label{eq:b3}
\end{eqnarray}
where $N_{m-1}^2 = f^{m}(\alpha_0)-\alpha_0$. 

When the functional $f(J_{0})$ is linear in $J_{0}$, i.e., 
$f(J_{0}) = q^2 J_0 +s$, it was shown in \cite{jpa} that the
algebra in eqs. (\ref{eq:alg1}-\ref{eq:alg3}) is a generalization
of $q$-oscillators, reducing to $q$-oscillators for $\alpha_0 = 0$.
Moreover, it was shown in \cite{jpa}, where the representation
theory was constructed in detail for the linear and quadratic
functions $f(x)$, that the essential tool in order to construct
representations of the algebra in (\ref{eq:alg1}-\ref{eq:alg3})
for a general analytic function $f(x)$ is the analysis of the
stability of the fixed points of $f(x)$ and their composed
functions.

It was shown in \cite{jpa} and \cite{comhugo} that there is a class
of one-dimensional quantum systems that are described by these
generalized Heisenberg algebras. This class is characterized by
those quantum systems having energy eigenvalues that can be
written as $\epsilon_{n+1} = f(\epsilon_{n})$,
where $\epsilon_{n+1}$ and $\epsilon_{n}$ are successive energy
levels and $f(x)$ a different function for each physical
system. This function $f(x)$ is exactly the same function that
appears in the construction of the algebra in eqs. 
(\ref{eq:alg1}-\ref{eq:alg3}). In this algebraic description 
of the class of quantum systems, $J_0$ is the Hamiltonian 
operator of the system, $A^{\dagger}$ and $A$ are the creation 
and annihilation operators that are related as in eq. 
(\ref{eq:casimir}) where $C$ is the Casimir operator of the 
representation associated to the quantum system.

\section{Relativistic square-well algebra}

We are going to construct in this section an algebraic formalism,
similar to the harmonic oscillator algebra, for the infinite
one-dimensional square-well potential in relativistic quantum
mechanics: the quantum one dimensional free Klein-Gordon
equation defined on the interval $0 \leq x \leq L$ with special 
boundary conditions. 
This Heisenberg-like algebra, that we call
relativistic square-well algebra, is an example in the large 
class of generalized Heisenberg algebras described in the 
previous section for a specific functional $f(x)$ that we
shall determine. In \cite{comhugo} it was constructed the
Heisenberg-like algebra of a non-relativistic particle in a
square-well potential, here in this section we present the
relativistic generalization of this algebra.

We briefly review the formalism of non-commutative differential 
and integral calculus on a one-dimensional lattice developed
in \cite{dimakis1} and \cite{dimakis2}. Let us consider an  
one dimensional lattice in a momentum 
space where the 
momenta are allowed only to take discrete values, say $p_{0}$, 
$p_{0}+a$, $p_{0}+2a$, $p_{0}+3a$ etc, with $a>0$.

The non-commutative differential calculus is based on the 
expression$^{\cite{dimakis1}, \cite{dimakis2}}$ 
\begin{equation}
    [p,dp] = dp \, a
    \label{eq:noncom1} \, ,
\end{equation}
implying that 
\begin{equation}
    f(p) \, dg(p) = dg(p) \, f(p+a) \, ,
    \label{eq:noncom2}
\end{equation}
for all functions $f$ and $g$. We introduce partial 
derivatives by 
\begin{equation}
    d \, f(p) = dp \, (\partial_{p} \, f) \, (p) = 
    (\bar{\partial}_{p} \, f) \, (p) \, dp \, ,
    \label{partial}
\end{equation}    
where the left and right discrete  derivatives are given by 
\begin{eqnarray}
    (\partial_{p} \, f) \, (p) & = & \frac{1}{a} \, [f(p+a) - f(p)] \, ,
    \label{eq:partialleft}  \\
   (\bar{\partial}_{p} \, f) \, (p)  & = & \frac{1}{a} \, [f(p) - f(p-a)] \, , 
    \label{eq:partialright}
\end{eqnarray}
that are the two possible definitions of derivatives on a lattice.
The Leibniz rule for the left discrete derivative can be written as,
\begin{equation}
    (\partial_{p} \, fg) \, (p) = (\partial_{p}f) \, (p) g \, (p) + 
    f(p+a)(\partial_{p} g) \, (p) \, ,
    \label{eq:leibniz}
\end{equation}
with a similar formula for the right derivative$^{\cite{dimakis1}}$.

Let us now introduce the momentum shift operators 
\begin{eqnarray}
    T  & = & 1 + a \, \partial_{p}
    \label{eq:a}  \\
    \bar{T} & = & 1 - a \, \bar{\partial}_{p} \, ,
    \label{eq:abarra}
\end{eqnarray}
which increases (decreases) the value of the momentum by $a$
\begin{eqnarray}
    (Tf) \, (p) & = & f(p+a)
    \label{eq:af}  \\
    (\bar{T}f) \, (p) & = & f(p-a)
    \label{eq:abarraf}
\end{eqnarray}
and satisfies 
\begin{equation}
    T \, \bar{T} = \bar{T} T = \hat{1} \, ,
    \label{eq:aabarra}
\end{equation}
where $\hat{1}$ means the identity on the algebra of functions of $p$.  

Introducing the momentum operator $P$$^{\cite{dimakis1}}$
\begin{equation}
    (Pf) \, (p) = p \, f(p) \, ,
    \label{eq:momentum}
\end{equation}
we have 
\begin{eqnarray}
    T P & = & (P+a)T
    \label{eq:ap}  \\
    \bar{T} P & = & (P-a) \bar{T} \, \, .
    \label{eq:abarrap}
\end{eqnarray}

Integrals can also be defined in this formalism. It is shown
in ref. \cite{dimakis1} that the property of an indefinite
integral 
\begin{equation}
\int df = f + \mbox{periodic function in} \, \, a \,\, ,
    \label{eq:intindef}
\end{equation}
suffices to calculate the indefinite integral of an
arbitrary one form. It can be shown that$^{\cite{dimakis1}}$
for an arbitrary function $f$
\begin{equation}
\int d\bar{p} \, f(\bar{p}) = \left\{
\begin{array}{lll}
a \sum_{k=1}^{[p/a]} f(p-k a) \, , & \mbox{if $p \geq a$} \\
0 \, \, ,  & \mbox{if $0 \leq p < a$} \\
-a \sum^{-[p/a]-1}_{k=0} f(p+k a) \, ,  & \mbox{if $p < 0$}
\end{array}
\right.
\label{eq:intresult}
\end{equation}
where $[p/a]$ is by definition the highest integer $\leq p/a $.

All equalities involving indefinite integrals are understood
modulo the addition of an arbitrary function periodic in $a$.
The corresponding definite integral is well-defined when the 
length of the interval is multiple of $a$. Consider the integral
of a function $f$ from $p_d$ to $p_u$ ($p_u = p_d+M a$, where
$M$ is a positive integer) as
\begin{equation}
\int_{p_d}^{p_u} dp f(p) = a \sum_{k=0}^{M} f(p_d+k a).
    \label{eq:intdef}
\end{equation}
Using eq. (\ref{eq:intdef}), 
an inner product of two (complex) functions 
$f$ and $g$ can be defined as 
\begin{equation}
    \langle f \, , \, g \rangle = \int_{p_{d}}^{p_{u}} dp \, f(p)^{*} \, g(p) \, , 
    \label{eq:inner}
\end{equation}
where $^{*}$ indicates the complex conjugation of the function $f$.  
The norm $\langle f \, , \, f \rangle \geq 0$ is
zero only when $f$ is identically null.  The set of equivalence classes  
\footnote{Two functions are in the same equivalence class if their
values coincide on all lattice sites.}
of normalizable functions $f$ ($\langle f \, , \, f \rangle $ is finite) 
is a Hilbert space.  
It can be shown that$^{\cite{dimakis1}}$
\begin{equation}
    \langle f, T g \rangle = \langle \bar{T} f, g \rangle \, ,
    \label{eq:inner}
\end{equation}
so that  
\begin{equation}
    \bar{T} = T^\dagger \, \, ,
    \label{eq:adjoint}
\end{equation}
where $T^\dagger$ is the adjoint operator of $T$.  Eqs. (\ref{eq:aabarra}) and 
(\ref{eq:adjoint}) show that $T$ is a unitary operator. Moreover, it is easy
to see that $P$ defined in eq. (\ref{eq:momentum}) is an Hermitian operator
and from (\ref{eq:adjoint}) one has
\begin{equation}
(i \partial_p)^{\dagger} = i \bar{\partial}_p  \, \, .
\end{equation} 

Now, we go back to address our main problem of this section, i.e. to 
construct an algebraic formalism, similar to the harmonic
oscillator algebra, for the infinite one-dimensional square-well 
potential in relativistic quantum mechanics. Let us assume we
have the one dimensional quantum free Klein-Gordon equation 
defined on the interval $0 \leq x \leq L$ 
with $\phi(x = 0,t)=\phi(x = L,t)=0$ where $\phi(x,t)$
is the one dimensional Klein-Gordon field.
It can easily be checked that the solution of this equation
is very similar to the solution of the Heisenberg equation
for the square-well potential.
The stationary part of the solution can be interpreted as 
being the spectrum of the 
Hamiltonian $H= \sqrt{P^{\, 2}+m_q^2}$,  $\hbar = c =1$, 
where the momentum is quantized with eigenvalues
$n \pi /L$ for $n = 1, 2, 3, \ldots$. Therefore, 
the momentum space is an one-dimensional periodic lattice 
with constant spacing $a= \pi/L$, clearly 
a candidate to apply the non-commutative differential 
calculus sketched before. We then take the momentum operator, $P$,
in the Hamiltonian $H= \sqrt{P^{\, 2}+m_q^2}$
as defined in eq. (\ref{eq:momentum}).

We can rewrite the Hamiltonian's eigenvalue associated with the 
(n+1)-th level as
\begin{equation}
    e_{n+1}^2 =  ( \sqrt{e_{n}^2 -m_q^2} + a )^2 + m_q^2  \, ,
    \label{eq:recorrencia}
\end{equation}
where $e_{n}$ is the Hamiltonian's eigenvalue associated with 
the $n$-th level and $a= \pi/L$ is the lattice spacing.

As $J_{0}$ is related to the 
Hamiltonian$^{\cite{algebra1}}$ and their eigenvalues 
are the iterations given 
by a function $f$ , we see that 
if we choose this function as
\begin{equation}
    f(x) = \sqrt{\left( \sqrt{x^2-m_q^2} + a \right)^2 + m_q^2} \, ,  
    \label{eq:deff}
\end{equation}
the $J_0$ in eqs. (\ref{eq:b1}-\ref{eq:b3}) has eigenvalues
equal to the energy eigenvalues of the square-well potential. 
Note that $\alpha_0 = \sqrt{a^2+m_q^2}$. Eqs. 
(\ref{eq:alg1}-\ref{eq:alg3}) can then be rewritten 
for this case as 
\begin{eqnarray}
 J_{0} A^{\dagger}  &=& A^{\dagger} 
\sqrt{\left( \sqrt{J_{0}^2-m_q^2} + a \right)^2 + m_q^2} \, , 
\label{eq:j+sw} \\
A J_{0} &=& \sqrt{\left( \sqrt{J_{0}^2-m_q^2} + a \right)^2 + m_q^2} \, \, 
\, \, \, \,A \, , 
\label{eq:j-sw} \\
\left[ A^{\dagger},A \right] &=& J_{0} - 
\sqrt{\left( \sqrt{J_{0}^2-m_q^2} + a \right)^2 + m_q^2} \, .
\label{eq:j0sw}
\end{eqnarray}
As $J_0$ is a Hermitian operator it can be diagonalized and, 
as we are considering
only the representations where the eigenvalues of $J_0$ are positive greater or equal to $m_q$,
the square roots in eqs. (\ref{eq:j+sw}-\ref{eq:j0sw}) are well defined.

We then have an algebra given in eqs. (\ref{eq:j+sw}-\ref{eq:j0sw}) where
the eigenvalues of $J_0$ , $e_n$, are the energy
eigenvalues of the relativistic one dimensional infinite
square-well potential and $A^{\dagger}$($A$) act as ladder operators. 
>From the previous discussion and from eqs. (\ref{eq:b1}-\ref{eq:b3}), taking
$\alpha_0 = \sqrt{a^2+m_q^2}$, the eigenvalues of $J_0$ when applied to the states
$| m \rangle$ give $f^m (m_q)$ that are, for $f$ given in eq. (\ref{eq:deff}),
the energy eigenvalues of the relativistic one dimensional infinite
square-well potential. Moreover, as said before and as seen from eqs.
(\ref{eq:b1}-\ref{eq:b3}), $A^{\dagger}$($A$) act as ladder operators.

In order
to have a complete description, similar to the case of the one-dimensional
harmonic oscillator, we must realize the operators $J_0$,
$A^{\dagger}$ and $A$ in
terms of the physical operators
of the system. The solution to this problem is 
\begin{eqnarray}
    A^{\dagger} & = &  S \, \bar{T} 
    \label{eq:j+}  \\
     A & = & T \, S    
    \label{eq:j-}  \\
    J_{0} & = & \sqrt{P^2 + m_q^2} \, .
    \label{eq:j0}
\end{eqnarray}
where $T$ and $\bar{T}$ are given by eqs. (\ref{eq:a}-\ref{eq:abarra}),
$P$ by eq. (\ref{eq:momentum}) and $S = (J_0 + C)^{1/2}$, with $C$
the Casimir of the algebra shown in eq. (\ref{eq:casimir}) having
eigenvalue $-\sqrt{a^2+m_q^2}$. 
Using eqs. (\ref{eq:ap}-\ref{eq:abarrap})
it is straightforward to check that the operators given in
eqs. (\ref{eq:j+}-\ref{eq:j0}) indeed satisfy eqs. 
(\ref{eq:j+sw}-\ref{eq:j0sw}). Moreover, using eq. (\ref{eq:aabarra})
we have
\begin{equation}
    A^{\dagger} \, A = S^2 =  \sqrt{P^2 + m_q^2} + C \, .
    \label{eq:joH}
\end{equation}

In summary, we have constructed an algebraic formalism,
similar to the harmonic oscillator algebra, for the
relativistic one-dimensional square-well potential. This
Heisenberg-like algebra is an example in the recently
constructed class of generalized Heisenberg algebras$^{\cite{jpa}}$. 
This class of algebras
contains also $q$-oscillators as an special case. Moreover,
it is also interesting to stress that the ladder operators
of the relativistic square-well algebra are realized in terms
of the physical operators of the system.

\section{Square-well quantum field theory} 

We are going to construct in this section a free
quantum field theory based on the relativistic
square-well algebra presented in the last section, i.e., the
vacuum excitations of this quantum field are particles
confined by the square-well potential.

In the momentum space appropriated to the construction of the
square-well algebra, as presented in the previous section,
besides the operator $P$ defined in eq. (\ref{eq:momentum})
one can define two type-coordinate self-adjoint operators as
\begin{eqnarray}
    \chi & = &  i (\bar{\partial}_p + \partial_p) \, \, ,
    \label{eq:cord1}  \\
     Q & = &  \bar{\partial}_{p} - \partial_{p} \, \, ,
    \label{eq:cord2}  
\end{eqnarray}
where $\partial_p$ and $\bar{\partial}_p$ are the left and right 
discrete derivatives defined in eqs. (\ref{eq:partialleft}, 
\ref{eq:partialright}). Of course, in the continuous limit
the operator $Q$ is identically null since $\partial_p$ and 
$\bar{\partial}_p$ represent the same derivative in this limit.

It can be checked that the operators $P$, $\chi$ and $Q$ generate
an algebra on the momentum lattice with lattice spacing
$a= \pi /L$$^{\cite{dimakis1}}$
\begin{eqnarray}
\left[ \chi,P \right] &=& 2 i \left( 1-\frac{a}{2} Q \right) ,
\label{eq:fecho1} \\
\left[ P,Q \right] &=& -i a \chi , 
\label{eq:fecho2} \\
\left[ \chi,Q \right] &=& 0 .
\label{eq:fecho3}
\end{eqnarray}
Note that, in the continuous limit $a \rightarrow 0$ we recover the standard
Heisenberg algebra, $\left[ x,p \right] = i$.

With the help of eqs. (\ref{eq:a}-\ref{eq:abarra} and \ref{eq:j+}-\ref{eq:j0}) 
we can rewrite $\chi$ and $Q$ as
\begin{eqnarray}
\chi &=&  \frac{i}{a} \left( S^{-1} A^{\dagger} - A S^{-1} \right)  ,
\label{eq:cord3} \\
Q &=&  \frac{1}{a} \left( -2 + S^{-1} A^{\dagger} + A \, S^{-1} \right) , 
\label{eq:cord4} 
\end{eqnarray}
where $S = (J_0 +C )^{1/2} = \left( \sqrt{P^2 + m_q^2} + C \right)^{1/2}$
with $C$ the Casimir operator shown in eq. (\ref{eq:casimir}) having
eigenvalue $-\sqrt{a^2+m_q^2}$ for the representation of interest. We stress
that $A^{\dagger}$ and $A$ are the creation and annihilation 
operators respectively, of a relativistic particle with mass $m_q$ in 
an infinite square-well potential as explained in the previous section.

Let us now introduce a three-dimensional discrete $\vec{k}$-space,
\begin{equation}
k_i = \frac{2 \pi l_i}{L_i} ,\, \,\,\,  i=1,2,3 \,\,\,\,\,\,\, ,
\label{eq:kspace}
\end{equation}
with $l_i= 0,\pm 1,\pm 2, \cdots $ and $L_i$, the lengths of the 
three sides of a rectangular box $\Omega$. For each point 
of this $\vec{k}$-space we associate
an independent copy of the one-dimensional
momentum lattice defined in the previous section  
such that $P_{\vec{k}}^{\dagger} = P_{\vec{k}}$ and 
$T_{\vec{k}}$ and $\bar{T}_{\vec{k}}$ 
are defined by means of the previous definitions,
eqs. (\ref{eq:a}-\ref{eq:abarra}),
through the substitution $P \rightarrow P_{\vec{k}}$
and $S_{\vec{k}}$ is given by
\begin{equation}
S_{\vec{k}} \equiv (J_0(\vec{k}) + C(\vec{k}))^{1/2} = 
\left( \sqrt{P^2_{\vec{k}} +m_q^2 +{\vec{k}}^2} + C(\vec{k}) \right)^{1/2} \, \, ,
\label{eq:defS}
\end{equation}
where $C(\vec{k})$ has for this representation eigenvalue $-\sqrt{a^2+m_q^2 + 
{\vec{k}}^2}$.

We now introduce for each point of this $\vec{k}$-space
independent operators $A^{\dagger}_{\vec{k}}$, 
$A_{\vec{k}}$ and $J_0(\vec{k})$ that commute for any
two different point of this $\vec{k}$-space and for the same
point we have
\begin{eqnarray}
 J_{0}(\vec{k}) A^{\dagger}_{\vec{k}}  &=& A^{\dagger}_{\vec{k}} 
\sqrt{\left( \sqrt{J_{0}^2(\vec{k})-m_k^2} + a \right)^2 + m_k^2} \, , 
\label{eq:j+swk} \\
A_{\vec{k}} J_{0}(\vec{k}) &=& \sqrt{\left( \sqrt{J_{0}^2(\vec{k})-m_k^2} + a \right)^2 + m_k^2} \, \, 
\, \, \, \,A_{\vec{k}} \, , 
\label{eq:j-swk} \\
\left[ A^{\dagger}_{\vec{k}},A_{\vec{k}} \right] &=& J_{0}(\vec{k}) - 
\sqrt{\left( \sqrt{J_{0}(\vec{k})^2-m_k^2} + a \right)^2 + m_k^2} \, ,
\label{eq:j0swk}
\end{eqnarray}
where $m_k = \sqrt{m_q^2 + \vec{k}^2}$. Since $J_0(\vec{k})$ is Hermitian
it can be diagonalized, moreover because the square roots the above algebra,
eqs. (\ref{eq:j+swk}-\ref{eq:j0swk}), are well-defined for representations 
where $J_0(\vec{k})$
eigenvalues are greater or equal to $m_k$.

It is not difficult to see that the realization of the
generators in terms of the physical operators is given as
\begin{eqnarray}
    A^{\dagger}_{\vec{k}} & = &  S_{\vec{k}} \, \bar{T}_{\vec{k}} \, ,
    \label{eq:j++k}  \\
     A_{\vec{k}} & = & T_{\vec{k}} \, S_{\vec{k}} \, ,    
    \label{eq:j--k}  \\
    J_{0}(\vec{k}) & = & \sqrt{P^2_{\vec{k}} + m_k^2} \, ,
    \label{eq:j0k}
\end{eqnarray}
where $T_{\vec{k}}$ and $\bar{T}_{\vec{k}}$ are given by eqs. 
(\ref{eq:a}-\ref{eq:abarra})
for $P \rightarrow P_{\vec{k}}$.

Now, we define the type-coordinate operators for
each point of the three-dimensional lattice as
\begin{eqnarray}
    \chi_{\vec{k}} & = &  i (\bar{\partial}_{p_{\vec{k}}} + 
\partial_{p_{-\vec{k}}}) \, \, ,
    \label{eq:cord5}  \\
     Q_{\vec{k}} & = &  \bar{\partial}_{p_{\vec{k}}} - 
\partial_{p_{-\vec{k}}} \, \, ,
    \label{eq:cord6}  
\end{eqnarray}
such that $\chi_{\vec{k}}^{\dagger}= \chi_{-\vec{k}}$ and 
$Q_{\vec{k}}^{\dagger}= Q_{-\vec{k}}$, exactly as it happens
in the construction of a spin-$0$ field for the spin-$0$ quantum 
field theory $^{\cite{tdlee}}$. With the previous definitions,
eqs. (\ref{eq:j++k}-\ref{eq:j--k} and \ref{eq:cord5}-\ref{eq:cord6}),
we can rewrite the type-coordinate operators in terms of the 
ladder operators of the square-well Heisenberg algebra
\begin{eqnarray}
\chi_{\vec{k}} &=&  \frac{i}{a} \left( -S^{-1}_{-\vec{k}} 
A^{\dagger}_{-\vec{k}} + A_{\vec{k}} S^{-1}_{\vec{k}} \right) \, ,
\label{eq:cord7} \\
Q_{\vec{k}} &=&  \frac{1}{a} \left( -2 + 
S^{-1}_{-\vec{k}} A^{\dagger}_{-\vec{k}} + 
A_{\vec{k}} \, S^{-1}_{\vec{k}} \right) \, . 
\label{eq:cord8} 
\end{eqnarray}

By means of the type-coordinate operators,
$\chi_{\vec{k}}S_{\vec{k}}$ and $Q_{\vec{k}}S_{\vec{k}}$, 
we can define two 
fields $\phi_1(\vec{r},t)$ and $\phi_2(\vec{r},t)$ as
\begin{eqnarray}
\phi_1(\vec{r},t) &=& \frac{1}{2} \sum_{\vec{k}} 
\frac{i}{\sqrt{\Omega}} \frac{1}{\omega(\vec{k})} \left( 
-S^{-1}_{\vec{k}} A^{\dagger}_{\vec{k}} S_{-\vec{k}}\, e^{-i \vec{k}.\vec{r}} + 
A_{\vec{k}} \, e^{i \vec{k}.\vec{r}} \right)  ,
\label{eq:defcampo1} \\
\phi_2(\vec{r},t) &=& \frac{1}{2} \sum_{\vec{k}} 
\frac{1}{\sqrt{\Omega}} \frac{1}{\omega(\vec{k})} \left( 
-2 S_{\vec{k}}\, e^{i \vec{k}.\vec{r}} + 
S^{-1}_{\vec{k}} A^{\dagger}_{\vec{k}} \, S_{-\vec{k}}\, e^{-i \vec{k}.\vec{r}}+ 
A_{\vec{k}} \, e^{i \vec{k}.\vec{r}} \right) , 
\label{eq:defcampo2} 
\end{eqnarray}
where $\omega(\vec{k})= \sqrt{\vec{k}^2+m^2}$, $m$ a real parameter
and $\Omega$ is the volume of a rectangular box.

What we have done so far is similar to the construction of spin-
$0$  fields in relativistic quantum field theory
in terms of the creation and annihilation operators
of the harmonic oscillator algebra
with the difference that
now, we have the ladder operators of a relativistic particle in
a square-well potential. Moreover, if we define
type-momentum fields as
\begin{eqnarray}
\Pi(\vec{r},t) &=&  \sum_{\vec{k}} \frac{\beta_1}{\sqrt{\Omega}}  
\, \, \, S_{\vec{k}}  \, e^{i \vec{k}.\vec{r}}   ,
\label{eq:defcampo3} \\
\wp(\vec{r},t) &=& \frac{1}{2\beta_1} \sum_{\vec{k}} 
\frac{1}{\sqrt{\Omega}}  \left( 
-S_{\vec{k}}\, e^{i \vec{k}.\vec{r}} + 
S^{-1}_{\vec{k}} A^{\dagger}_{\vec{k}} \, S_{-\vec{k}}\, 
e^{-i \vec{k}.\vec{r}}+ 
A_{\vec{k}} \, e^{i \vec{k}.\vec{r}} \right) , 
\label{eq:defcampo4} 
\end{eqnarray}
where $\beta_1$ is an arbitrary real number
we can show that the Hamiltonian
\begin{eqnarray}
H =  &\int& d^3 r \left(  \Pi(\vec{r},t)^{\dagger}
\, \, \wp(\vec{r},t) + \wp(\vec{r},t)^{\dagger} 
\, \, \Pi(\vec{r},t) + \right.  \\
 &&\left.   \phi_1(\vec{r},t)^{\dagger}  (-{\vec{\nabla}}^2+m^2)  
\phi_1(\vec{r},t) + \phi_2(\vec{r},t)^{\dagger}  (-{\vec{\nabla}}^2+m^2)  
\phi_2(\vec{r},t)  \right) \nonumber \,\, ,
\label{eq:defhamilt}
\end{eqnarray}
can be rewritten as
\begin{equation}
H = \sum_{\vec{k}} A^{\dagger}_{\vec{k}} A_{\vec{k}} =
\sum_{\vec{k}} S_{\vec{k}}^2 = \sum_{\vec{k}} 
\sqrt{P_{\vec{k}}^2 + m_q^2 + {\vec{k}}^2} + C(\vec{k}) \,\, ,
\label{eq:resulthamilt}
\end{equation}
where $P_{\vec{k}}$, for each $\vec{k}$, is the momentum operator for a
particle with mass $m_k$ in a square-well potential. The Casimir operator,
$C(\vec{k})$, in the representation in consideration has eigenvalue
$-\sqrt{a^2+m_q^2 + {\vec{k}}^2}$.

The eigenvectors of $H$ form a complete set and span the Hilbert
space of this system. The eigenvectors are
\begin{equation} 
|1 \rangle, \,\, A^{\dagger}_{\vec{k}} |1 \rangle, \,\,
A^{\dagger}_{\vec{k}} A^{\dagger}_{\vec{k}'} |1 \rangle \,\,
\mbox{for} \,\, \vec{k}\not= \vec{k}', \,\,
(A^{\dagger}_{\vec{k}})^2 |1 \rangle, \,\, \cdots
\label{eq:hilbert} 
\end{equation}
Note that the lowest energy state is the one of a particle in the lowest
level in the relativistic square-well potential which is
represented by $|1 \rangle$.
This Hilbert space has a different interpretation with
respect to the standard spin-$0$ quantum field theory based
on the harmonic oscillator. In the standard case the analog 
of the formula (\ref{eq:resulthamilt}) is
\begin{equation}
H = \sum_{\vec{k}}  N_{\vec{k}} \, \sqrt{\vec{k}^2+m^2} \,\, ,
\label{eq:standardresulthamilt}
\end{equation}  
where $N$ is the number operator. While in the standard
quantum field theory the creation operator creates one
particle of mass $m$ each time it is applied to the vacuum, 
in the 
square-well quantum field theory one reads from eq. 
(\ref{eq:resulthamilt}) that the creation operator in this
case creates excited states of particles with mass spectrum
given by $\sqrt{a^2 \, n^2 +m_q^2}$ where 
$n = 1,2,\cdots$ 
is the level of the particle with mass $m_q$ in the square-well
confining potential. In other words, for instance, the state 
$(A_{\vec{k}}^{\dagger})^n \, |1 \rangle$ represents not a $n$-particle state
but a state of one particle with mass $\sqrt{a^2 \, (n+1)^2 +m_q^2}$.

The time evolution of the fields can be studied by means
of Heisenberg's equation for $A^{\dagger}_{\vec{k}}$ and 
$A_{\vec{k}}$. Taking into account
eqs. (\ref{eq:j+sw}-\ref{eq:j0sw}, \ref{eq:resulthamilt}) 
we have
\begin{equation}
\frac{d}{dt} A^{\dagger}_{\vec{k}}(t) \equiv
{\dot{A}}^{\dagger}_{\vec{k}}(t) = i A^{\dagger}_{\vec{k}}(t) \,\,
\Delta H_{\vec{k}} \,\,\, ,
\label{eq:evolution}
\end{equation}
with
\begin{equation}
\Delta H_{\vec{k}} = \sqrt{(P_{\vec{k}}+a)^2 + m_q^2 + {\vec{k}}^2}
- \sqrt{P_{\vec{k}}^2 + m_q^2 + {\vec{k}}^2} \,\,\, .
\label{eq:deltah}
\end{equation}
Eq. (\ref{eq:evolution}) has as solution
\begin{equation}
A^{\dagger}_{\vec{k}}(t) =  A^{\dagger}_{\vec{k}}(0)  \,
\exp\left(i \Delta H_{\vec{k}} t\right) \,\,\, ,
\label{eq:solution}
\end{equation}
with a similar expression for $A_{\vec{k}}(t)$.

Even for the 
free theory we have just constructed
there are some additional points that deserve to be further investigated.
For example, it would be interesting
to construct the Lorentz algebra operators of the system.
We hope to understand this and other points in a near future.

\section{Final comments}

This paper was the first step towards
an analysis of possible consequences in
QFT of a class of 
generalized Heisenberg algebras we have recently
constructed $^{\cite{jpa}}$. We hope that the QFT we developed
here can be used as an alternative 
phenomenological approach to hadronic interactions and also
that the entire treatment developed here could be applied
to a deformed Heisenberg algebra in order to construct
a deformed QFT for very high energy physics 
($10^3$Gev$ < E < 10^{19}$Gev).

We constructed here a free quantum 
field theory based on the Heisenberg algebra of a relativistic particle
in an infinite square-well potential: the quantum one dimensional free
Klein-Gordon equation defined on the interval
$0 \leq x \leq L$ with special boundary conditions. 
It is interesting to stress that this formalism shows that it is 
possible to construct a different class of quantum field theories
based on an algebraic structure, having ladder operators,
different from the harmonic oscillator algebra.
Moreover, note that even if the Heisenberg relativistic 
square-well algebra shown in eqs. (\ref{eq:j+sw}-\ref{eq:j0sw}) has
a non-simple form, the realization of the ladder operators in terms
of the physical operators of the model as seen in eqs. 
(\ref{eq:a}-\ref{eq:abarra} and \ref{eq:j+}-\ref{eq:j0}) 
is really very simple. 

In the previous section, we introduced fields written in terms
of the generators of the square-well algebra and a 
Hamiltonian written in terms of these fields describing a system
with an infinite number of degrees of freedom and mass spectrum 
excitations given by $\sqrt{n^2 \pi^2/L^2+m_q^2}$,
where $n = 1,2,\cdots$
denotes the level of the particle with mass $m_q$ in an infinite
square-well potential of width $L$. As we have already noted, even 
for this free quantum field theory 
there are certainly some points, not analyzed in this
paper, as the mentioned Lorentz algebra, that deserve to be investigated.
It would be also interesting to
construct the interacting theory of this quantum field theory,
for the excitations produced by the square-well creation operator
as well as for their interaction with an electromagnetic field.

Finally, since the mass spectrum of this quantum field theory
are excited states of a confined particle it is tempting
to use it 
to try to address at least some aspects of hadronic interactions.
In other words, we conjecture a possible application of this
formalism as a phenomenological quantum field approach to hadronic 
interactions. 
Of course, as the "potential" (or the mass spectrum) used in this
quantum field theory is the simple square-well potential it is
hard to imagine that the results given by the interacting quantum field 
theory can be very precise. 
Thus, we also think it would be interesting to extend the formalism 
developed in this paper to a more realistic mass spectrum.

\vspace{0.7 cm}
\noindent
{\bf Acknowledgments:} The authors thank L. Castellani, S. Sciuto and 
F. R. A. Sim\~{a}o for comments and PRONEX for 
partial support. M. A. R. M.  thanks the members of the 
INFN sezione di Torino for their kind hospitality.
E. M. F. C. thanks CNPq for a grant.

\newpage

\end{document}